\documentclass{ws-procs9x6-rev}            
\usepackage{slashed}
\begin{document}
\vspace*{-1cm}
\begin{flushright}
DESY 16-090\\May 2016
\end{flushright}

\title{Multiquark Hadrons - A New Facet of QCD\\}

\author{Ahmed Ali $^*$}

\address{Deutsches Elektronen-Synchrotron DESY,\\
D-22607 Hamburg, Germany\\
$^*$E-mail: ahmed.ali@desy.de\\}

\begin{abstract}
I review some selected aspects of the phenomenology of multiquark states discovered in high energy experiments.
They have four valence quarks (called tetraquarks) and two of them are found to have
five valence quarks (called pentaquarks), extending the conventional  hadron spectrum which consists
of quark-antiquark $(q\bar{q})$ mesons and $qqq$ baryons. Multiquark states  represent a new facet of QCD
and their dynamics is both challenging and  currently poorly understood.  I discuss various approaches put
forward to accommodate them, with  emphasis on the diquark model. 
\end{abstract}


\bodymatter

\section{Introduction}\label{ali:sec1}
Ever since the discovery of the state $X(3870)$ by Belle in 2003~\cite{Belle-C},
 a large number of multiquark states has been discovered in particle physics experiments
(see recent reviews~\cite{Chen:2016qju,Olsen:2015zcy,Briceno:2015rlt,Esposito:2014rxa,Bodwin:2013nua}). 
 Most of them are quarkonium-like states, in that they have a $(c\bar{c})$
or a $(b\bar{b})$ component in their Fock space. A good fraction of them  
 is electrically neutral but some are singly-charged.
 Examples are $X(3872) (J^{PC}=1^{++})$,  $Y(4260) (J^{PC}=1^{--})$, $Z(3900)^\pm (J^P=1^+)$,
$P_c(4450)^\pm (J^P=5/2^+)$,  in the hidden charm sector, and $Y_b(10890) (J^{PC}=1^{--})$, $Z_b(10610)^\pm
(J^P=1^+)$ and $Z_b(10650)^\pm (J^P=1^+)$, in the hidden bottom sector. The numbers in the parentheses are
their masses in MeV. Of these,  $P_c(4450)^\pm (J^P=5/2^+)$ is a pentaquark state, as its discovery mode
$P_c(4450)^+ \to J/\psi p$ requires a minimal valence quark content $c \bar{c} u u d$. The others are tetraquark
states, with characteristic decays, such as $X(3872) \to J/\psi \pi^+ \pi^-$, $Y(4260) \to J/\psi \pi^+ \pi^-$, 
$Z(3900)^+ \to J/\psi \pi^+$, $Y_b(10890) \to \Upsilon(1S,2S,3S) \pi^+ \pi^-$, and $Z_b(10610)^+ \to h_b(1P,2P) \pi^\pm,
\Upsilon(1S,2S) \pi^+$. No doubly-charged multiquark hadron has been seen so far, though some are expected, such as
$[\bar{c}\bar{u}][sd] \to D_s^- \pi^-$, in the tetraquark scenario discussed below.

Deciphering the underlying dynamics of the multiquark states is a formidable challenge and several models have
been proposed to accommodate them. They include, among others, cusps~\cite{Swanson:2015bsa,Swanson:2014tra},
which assume that the final state rescatterings  are enough to describe data, and as such
there is no need for poles in the scattering matrix. This is the minimalist approach, in particular, invoked
to explain the origin of the charged states $Z_c(3900)$ and $Z(4025)$. If proven correct, one would have to admit
that all this excitement about new frontiers of QCD  is ``much ado about nothing''.

 A good majority of the interested hadron physics community obviously does not share this agnostic point of
view, and dynamical mechanisms have been devised to accommodate the new spectroscopy.  One such model put forward to accommodate the exotic hadrons
is hadroquarkonium, in which a $Q\bar{Q}$ $(Q=c, b)$ forms the hard core surrounded
by light matter (light $q\bar{q}$ states). For example, the hadrocharmonium core may consist of $J/\psi, \psi^\prime, \chi_c$
states, and the light $q\bar{q}$ degrees of freedom can be combined to accommodate the observed
 hadrons~\cite{Dubynskiy:2008mq}.
 This is motivated by analogy with the good old hydrogen atom
which explained a lot of atomic physics. A variation on this theme is that the hard core quarkonium
could be in a color-adjoint representation, in which case the light degrees of freedom are also a
color-octet to form an overall singlet. 

Next are hybrid models, the basic idea of which dates back to circa 1994~\cite{Close:1994hc}  based on the QCD-inspired
flux-tubes, which predict
exotic $J^{\rm PC}$ states of both the light and heavy quarks. Hybrids are hadrons formed from the valence quarks
and gluons, for example, consisting of $q\bar{q}g$. 
 In the context of the $X,Y,Z$ hadrons, hybrids
have been advanced  as a model for the $J^{\rm PC}=1^{--}$ state $Y(4260)$, which has a small $e^+e^-$
annihilation cross section\cite{Close:2005iz,Kou:2005gt,Zhu:2005hp}, But, hybrids have been offered as
templates for other exotic hadrons as well~\cite{Meyer:2015eta,Pennington:2015pda}.

Another popular approach assumes that the multiquark  states are  meson-meson and  meson-baryon bound states,
with an attractive residual van der Waals force generated by mesonic exchanges~\cite{molecule-refs}.  
This hypothesis is in part supported by the closeness of the observed exotic hadron masses to the respective
 meson-meson (meson-baryon) thresholds. In many cases, this leads to very small binding energy, which imparts
 them a very large hadronic radius. This is best illustrated by  $X(3872)$, 
which has an S-wave coupling to $D^* \bar{D}$ (and its conjugate) and  has a binding energy 
 ${\cal E}_X=M_{X(3872)}-M_{D^{*0}} -M_{\bar{D}^0}=-0.3 \pm 0.4$ MeV.  Such a hadron molecule will have a large mean square separation of the constituents
$\langle r_X \rangle \propto 1/\sqrt{{\cal E}_X} \simeq 5$ fm, where the quoted radius corresponds to a binding energy
 ${\cal E}_X=0.3$ MeV. This would lead to small production cross-sections in hadronic collisions~\cite{Bignamini:2009sk}, contrary to
what has been observed in a number of experiments at the Tevatron and the LHC. In some theoretical constructs,
this problem is mitigated by making the hadron molecules complicated by invoking a
hard (point-like) core. In that sense, such models resemble hadroquarkonium models, discussed above.
In yet others, rescattering effects are invoked to substantially increase the cross-sections~\cite{Artoisenet:2009wk}.
Theoretical interest in hadron molecules has remained unabated, and 
 there exists a vast and growing literature on this topic with ever increasing sophistication, a sampling of which
is referenced here
~\cite{Gutsche:2014zda,Cleven:2014qka,Barnes:2014csa,Guo:2016bjq,Artoisenet:2010va,Hanhart:2011jz,Meng:2014ota}.

Last, but by no means least, on this list are QCD-based interpretations in which tetraquarks and pentaquarks
are genuinely new hadron species~\cite{Maiani:2004uc,Maiani:2004vq,Brodsky:2014xia}.
 In the large $N_c$ limit of QCD, tetraquarks 
 are shown to exist~\cite{Weinberg:2013cfa,Knecht:2013yqa,Rossi:2016szw} as poles in the S-matrix,
and they may have narrow widths in this approximation, and hence they are reasonable candidates for
 multiquark states.
First attempts using Lattice QCD have been undertaken~\cite{Padmanath:2015pkj,DeTar:2015orc} in which
correlations involving four-quark operators are studied numerically. 
Evidence of tetraquark states in the sense of  S-matrix poles using these methods
is still lacking. Establishing the signal of a resonance requires good control of the
background. In the lattice QCD simulations of multiquark states, this is currently not the case. This may be
traced back to the presence of a number of nearby hadronic thresholds and to lattice-specific issues,
such as an unrealistic pion mass. More powerful analytic and computational techniques are needed to
draw firm conclusions.
In the absence of reliable first principle calculations, approximate phenomenological methods are the
only way forward. In that spirit,
 an effective Hamiltonian approach has been often
 used~\cite{Maiani:2004uc,Maiani:2004vq,Maiani:2014aja,Ali:2009pi,Ali:2009es,Ali:2010pq}, 
in which tetraquarks are assumed to be  diquark-antidiquark objects, bound by gluonic exchanges
 (pentaquarks are diquark-diquark-antiquark objects). This allows one to work out the
spectroscopy and some aspects of tetraquark decays. Heavy quark symmetry is a help in that it can be used
for the heavy-light diquarks relating the charmonia-like states to the bottomonium-like counterparts.
I will be mainly discussing interpretations of the current data based on the phenomenological diquark picture to
test how far such models go in describing the observed exotic hadrons and other properties measured 
in current experiments.

\section{The Diquark Model}

The basic assumption of this  model is that diquarks are tightly bound colored objects and
they are the building blocks for forming tetraquark mesons and pentaquark baryons.
The diquarks, for which we use the notation $[qq]_c$, and interchangeably ${\mathcal Q}$,
have two possible SU(3)-color representations.
Since quarks transform as a triplet $\tt 3$ of color SU(3), the diquarks resulting from the
direct product $\tt 3 \otimes 3=\bar{3} \oplus 6$, are thus either a color anti-triplet $\tt \bar{3}$ or a
color sextet $\tt 6$. The leading diagram based on one-gluon exchange is
shown below.
\begin{figure}
\centerline{\includegraphics[width=5cm]{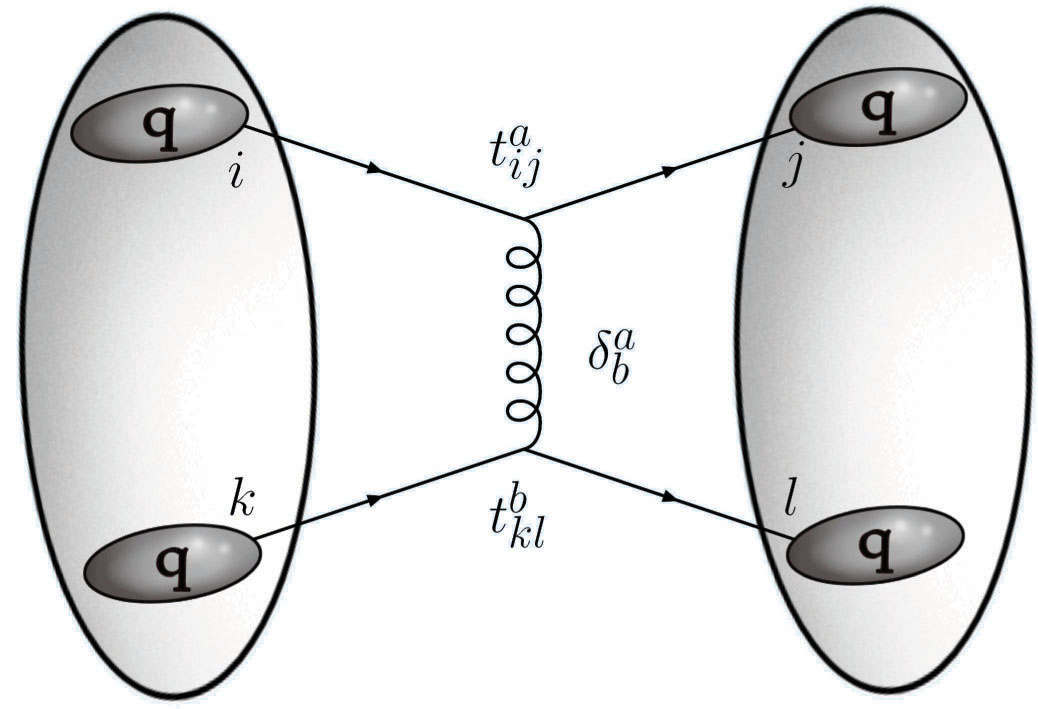}}
\caption{One-gluon exchange diagram for diquarks.}
\label{ali:fig1}
\end{figure}

\noindent The product of the SU(3)-matrices in Fig.~\ref{ali:fig1} can be decomposed as
\begin{equation}
\vspace{-0.3cm}
t^a_{ij}t^a_{kl}=
-\frac{2}{3}\underbrace{(\delta_{ij}\delta_{kl}-\delta_{il}\delta_{kj})/2}_{\rm{antisymmetric:\; projects}\enspace {  \bar{\mathbf 3} }}
+\frac{1}{3}\underbrace{(\delta_{ij}\delta_{kl}+\delta_{il}\delta_{kj})/2}_{\rm{symmetric:\; projects}\enspace { \mathbf 6}}~.
\nonumber
\end{equation}
\vspace*{3mm}

\noindent The coefficient of the antisymmetric $\tt \bar{3}$ representation is $-2/3$, reflecting that the two diquarks bind with a
strength half as strong as between a quark and an antiquark, in which case the corresponding coefficient is $-4/3$.
The symmetric $\tt 6$ on the other hand has a positive coefficient, +1/3,  reflecting a repulsion. This perturbative
 argument is in agreement with lattice QCD simulations~\cite{Alexandrou:2006cq}. Thus,  in working out the phenomenology,
 a diquark is assumed to be
an $SU(3)_c$-antitriplet, with the antidiquark a color-triplet. With this, we have two color-triplet fields, quark $q_3$
 and anti-diquark
 $\overline{\mathcal Q}$ or $[\bar{q}\bar{q}]_{3}$, 
and two color-antitriplet fields, antiquark $\bar{q}_{\bar{3}}$ and diquark  ${\mathcal Q}$ or $[qq]_{\bar{3}}$,
from which the spectroscopy of the conventional and exotic  hadrons is built.

\begin{figure}
\centerline{\includegraphics[width=8cm]{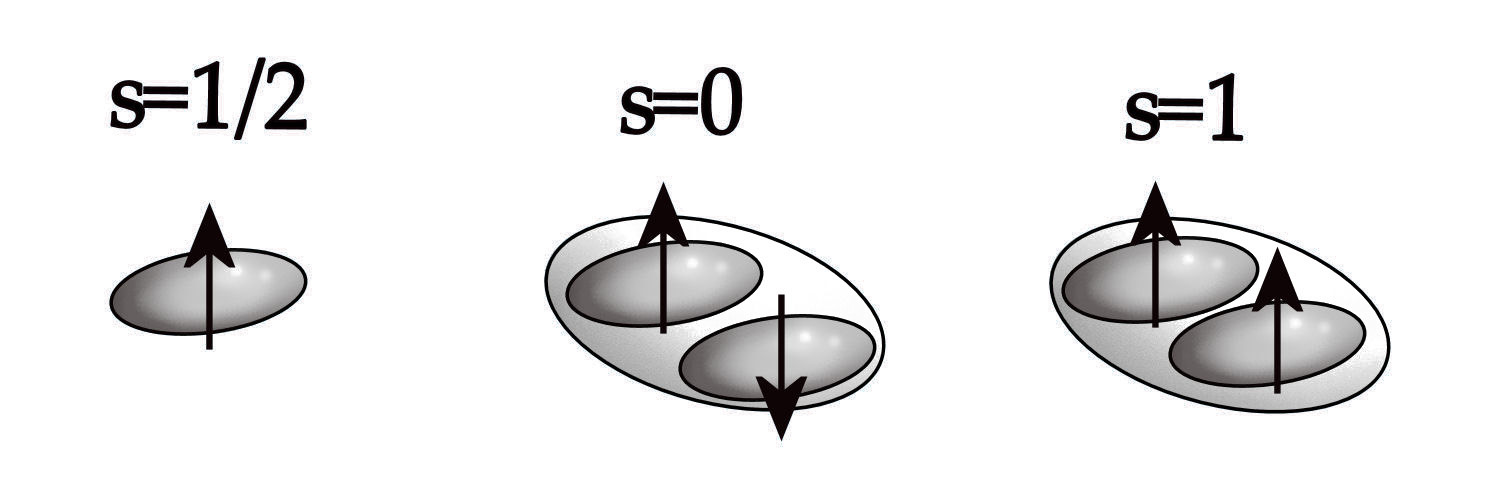}}
\caption{Quark and diquark spins.}
\label{ali:fig2}
\end{figure}

 Since quarks are spin-1/2 objects, a diquark has two possible spin-configurations, spin-0, with the
two quarks in a diquark having their spin-vectors anti-parallel, and spin-1, in which case the two quark spins
 are aligned, as shown in Fig.~\ref{ali:fig2}.
They were given the names ``good diquarks'' and ``bad diquarks'', respectively, by Jaffe~\cite{Jaffe:2004ph}, implying that in
 the former case, the two quarks bind, and in the latter, the binding is not as strong.
 There is some support of this
 pattern from lattice simulations for light diquarks~\cite{Alexandrou:2006cq}. However, as the spin-degree of freedom decouples in the heavy quark systems, as can be shown
explicitly in heavy quark effective theory context for heavy mesons and baryons, we expect that this decoupling
will also hold for heavy-light diquarks $[Q_i q_j]_{\bar{3}}$ with $Q_i=c, b; q_j=u,d,s$. So, for the heavy-light diquarks,
 both the spin-1 and spin-0 configurations are present. Also, what concerns the
diquarks in heavy baryons (such as $\Lambda_b$ and $\Omega_b$), consisting of a heavy quark and 
a light diquark,  both $j^p=0^+$ and $j^p=1^+$ quantum numbers of the diquark are needed to accommodate the observed baryon spectrum.

In this lecture, we will be mostly discussing heavy-light diquarks, and following the discussion above, we
construct the interpolating diquark operators for the two spin-states of such diquarks
 (here $Q=c,b$)~\cite{Maiani:2004vq}:
\vspace*{3mm}

%
\begin{tabular}{rlrcl}
$\textnormal{Scalar }$ &
$0^+$: &
$\mathcal Q_{i \alpha} $
&=& 
$\epsilon_{\alpha\beta\gamma}
(\bar{Q}_c^{\beta}\gamma_5
q_i^{\gamma}
-\bar{q}_{i_c}^\beta \gamma_5 Q^\gamma),\qquad_{_{\textnormal{$\alpha, \beta, \gamma$: $SU(3)_C$  indices} }}$
\\
$\textnormal{Axial-Vector}$&
$1^+ $:
&
$\vec{\mathcal Q}_{i \alpha} $
&=&
$\epsilon_{\alpha\beta\gamma}(\bar{Q}_c^{\beta} \vec{\gamma} q_i^{\gamma}
+
\bar{q}_{i_c}^\beta \vec{\gamma} Q^\gamma).
$
\end{tabular}
%
%
\vspace*{2mm}

\noindent In the non-relativistic (NR) limit, these states are parametrized by Pauli matrices:
$ \Gamma^0 =
\frac{\sigma_2}{\sqrt{2}} (\textnormal{Scalar}~0^+),$ 
and
$\;\;\vec{\Gamma} =
\frac{\sigma_2\vec{\sigma}}{\sqrt{2}} (\textnormal{Axial-Vector }1^+). 
$
We will characterize a tetraquark state with total angular momentum $J$ by the state vector
$\left\vert Y_{[bq]}\right\rangle=\left\vert s_{\mathcal Q},s_{\bar{\mathcal Q}};~J\right\rangle$
showing the diquark spin $s_{\mathcal Q}$ and the antidiquark spin $s_{\bar{\mathcal Q}}$.
Thus, the tetraquarks with the following diquark-spin and angular momentum $J$ have the Pauli forms: 
%
\begin{eqnarray}
\left\vert 0_{\mathcal Q},0_{\bar{\mathcal Q}};~0_{J}\right\rangle &=&\Gamma^0 \otimes \Gamma^0 ,  \notag \\
\left\vert 1_{\mathcal Q},1_{\bar{\mathcal Q}};~0_{J}\right\rangle &=&\frac{1}{\sqrt{3}}%
\Gamma^i \otimes \Gamma_i  \ldots,  \notag 
\\
\left\vert 0_{\mathcal Q},1_{\bar{\mathcal Q}};~1_{J}\right\rangle &=&\Gamma^0 \otimes \Gamma^i ,  \notag \\
\left\vert 1_{\mathcal Q},0_{\bar{\mathcal Q}};~1_{J}\right\rangle &=&\Gamma^i \otimes \Gamma^0 ,  \notag \\
\left\vert 1_{\mathcal Q},1_{\bar{\mathcal Q}};~1_{J}\right\rangle &=&\frac{1}{\sqrt{2}}%
\varepsilon ^{ijk}\Gamma_j \otimes \Gamma_k. \notag
\end{eqnarray}
 
\subsection{NR Hamiltonian for Tetraquarks with hidden charm}
For the heavy quarkonium-like exotic hadrons, we work in the non-relativistic limit and use
the following effective Hamiltonian to calculate the tetraquark mass
 spectrum~\cite{Maiani:2004vq,Maiani:2014aja}
$$
H_{\rm eff}=2m_{\mathcal Q}+H_{SS}^{(qq)}+H_{SS}^{(q\bar{q})}+H_{SL}+H_{LL}, 
$$
where $m_{\mathcal Q}$ is the diquark mass, the second term above is the spin-spin interaction involving
the quarks (or antiquarks) in a diquark (or anti-diquark), the third term depicts spin-spin interactions
involving a quark and an antiquark in two different shells, with the fourth and fifth terms being the
spin-orbit and the orbit-orbit interactions, involving the quantum numbers of the tetraquark, respectively.
For the $S$-states, these last two terms are absent. For illustration, we consider the case $Q=c$ and display
the individual terms in $H_{\rm eff}$:
\begin{eqnarray}
H_{SS}^{(qq)}= 2(\mathcal{K}_{cq})_{\bar{3}}[(\mathbf{S}_{c}\cdot \mathbf{S}_{q})
+(\mathbf{S}_{\bar{c}}\cdot \mathbf{S}_{\bar{q}})],
\notag \\
&\hspace{-5.8cm} H_{SS}^{(q\bar{q})}=2(\mathcal{K}_{c\bar{q}})(\mathbf{S}_{c}
\cdot \mathbf{S}_{\bar{q}}+\mathbf{S}_{\bar{c}}\cdot \mathbf{S}_{q})
 +2 \mathcal{K}_{c\bar{c}} (\mathbf{S}_{c}\cdot \mathbf{S}_{\bar{c}})
+2 \mathcal{K}_{q\bar{q}} (\mathbf{S}_{q}\cdot \mathbf{S}_{\bar{q}}),
\notag \\
&\hspace{-11.3cm} H_{SL}  = 2 A_{\mathcal Q} (\mathbf{S}_{\mathcal{Q}}\cdot \mathbf{L}+\mathbf{S}_{\mathcal{\bar{Q}} }\cdot \mathbf{L}), 
\notag \\
&\hspace{-12cm} H_{LL} = B_{\mathcal Q} \frac{L_{\mathcal Q\bar{\mathcal Q}}(L_{\mathcal Q\bar{\mathcal Q}}+1)}{2}.  
\notag 
\end{eqnarray}
\begin{figure}
\centerline{\includegraphics[width=7cm]{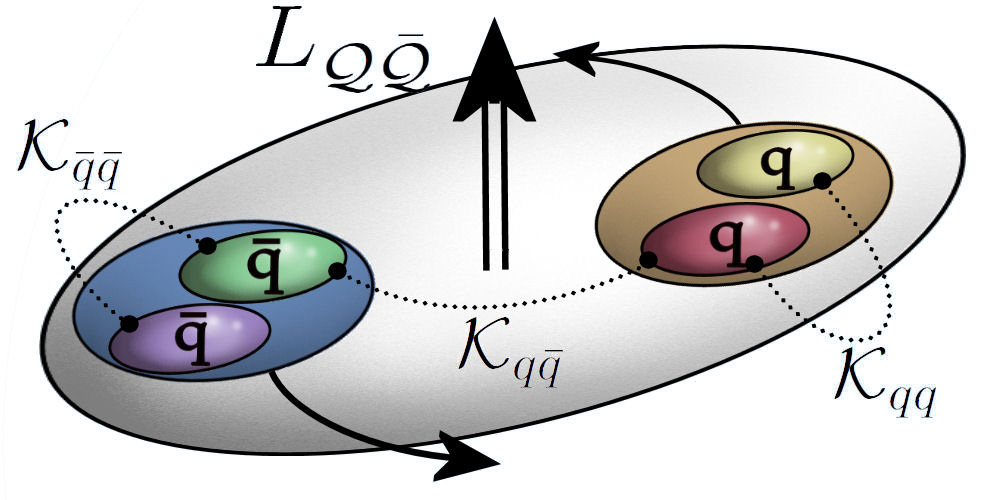}}
\vspace{-2mm}
\caption{Schematic diagram of a tetraquark in the diquark-antidiquark picture.}
\label{ali:fig3}
\end{figure}

\noindent The usual angular momentum algebra then yields the following form:
\begin{align}
H_{\rm eff}
&=2m_{\mathcal Q}
+ \frac{B_{\mathcal Q}}{2} \langle \bm{L}^2 \rangle
- 2a \langle \bm{L}\cdot \bm{S} \rangle
+ 2\kappa_{qc} \big[ \langle \bm{s_{q}}\cdot \bm{s_{c}} \rangle
 + \langle \bm{s_{\bar q}}\cdot \bm{s_{\bar c}} \rangle \big]
\nonumber\\
&=2m_{\mathcal Q}
- a J(J+1)
+ \bigg( \frac{B_{\mathcal Q}}{2} + a \bigg) L(L+1)
+ a S(S+1)- 3\kappa_{qc} \nonumber\\
&
+ \kappa_{qc}\big[ s_{qc}(s_{qc}+1) +
 s_{\bar{q}\bar{c}}(s_{\bar{q}\bar{c}}+1) \big].
\nonumber
\end{align}
\subsection{ Low-lying $S$ and $P$-wave tetraquark states in the $c\bar{c}$ and $b\bar{b}$ sectors}%

The states in the diquark-antidiquark basis $| s_{qQ}, s_{\bar{q}\bar{Q}}; S, L \rangle_J$
and in the $Q\bar{Q}$ and $q\bar{q}$ basis $| s_{q\bar{q}}, s_{Q\bar{Q}}; S', L' \rangle_J$ 
are related by Fierz transformation. The positive parity $S$-wave
tetraquarks are given in terms of the six
states listed in Table~\ref{ali:tbl1} (charge conjugation is defined for neutral states). These states
are characterized by the quantum number $L=0$, hence their masses depend on just two parameters
$M_{00}$ and $\kappa_{qQ}$, leading to  several predictions to be tested against experiments.
The $P$-wave states are listed in Table~\ref{ali:tbl2}. The first four of them have $L=1$, and the
fifth has $L=3$, and hence is expected to be significantly heavier.
\begin{table}
\begin{center}
\tbl{$S$-wave tetraquark states in two bases and their masses in the diquark model.}
{\begin{tabular}{ccccc}
\toprule
Label & $J^{PC}$ & 
$| s_{qQ}, s_{\bar{q}\bar{Q}}; S, L \rangle_J$ &
$| s_{q\bar{q}}, s_{Q\bar{Q}}; S', L' \rangle_J$ &
Mass
\\
\colrule
$X_0$ & $0^{++}$ & $| 0, 0; 0, 0 \rangle_0$ &
$\big( | 0, 0; 0, 0 \rangle_0 + \sqrt{3}| 1, 1; 0, 0 \rangle_0 \big)/2$ &
$M_{00} - 3\kappa_{qQ}$
\\
$X_0^\prime$ & $0^{++}$ & $| 1, 1; 0, 0 \rangle_0$ &
$\big( \sqrt{3}| 0, 0; 0, 0 \rangle_0 -| 1, 1; 0, 0 \rangle_0 \big)/2$ &
$M_{00} + \kappa_{qQ}$
\\
$X_1$ & $1^{++}$ &$\big( | 1, 0; 1, 0 \rangle_1 + | 0, 1; 1, 0 \rangle_1 \big)/\sqrt{2}$ &
$| 1, 1; 1, L' \rangle_1$ &
$M_{00} - \kappa_{qQ}$
\\
$Z$ & $1^{+-}$ & 
$\big( | 1, 0; 1, 0 \rangle_1 - | 0, 1; 1, 0 \rangle_1 \big)/\sqrt{2}$ &
$\big( | 1, 0; 1, L' \rangle_1 - | 0, 1; 1, L' \rangle_1 \big)/\sqrt{2}$ &
$M_{00} - \kappa_{qQ}$
\\
$Z^\prime$ & $1^{+-}$ & $| 1, 1; 1, 0 \rangle_1$ &
$\big( | 1, 0; 1, L' \rangle_1 + | 0, 1; 1, L' \rangle_1 \big)/\sqrt{2}$ &
$M_{00} + \kappa_{qQ}$
\\
$X_2$ &  $2^{++}$ & $| 1, 1; 2, 0 \rangle_2$ &
$| 1, 1; 2, L' \rangle_2$ &
$M_{00} + \kappa_{qQ}$
\\
\botrule
\end{tabular}
}
\label{ali:tbl1}
\end{center}
\end{table}
\begin{table}
\begin{center}
\tbl{$P$-wave tetraquark states in two bases and their masses in the diquark model.}
{\begin{tabular}{ccccc}
\toprule
Label & $J^{PC}$ & 
$| s_{qQ}, s_{\bar{q}\bar{Q}}; S, L \rangle_J$ &
$| s_{q\bar{q}}, s_{Q\bar{Q}}; S', L' \rangle_J$ &
Mass
\\
\colrule
$Y_1$ & $1^{--}$ & $| 0, 0; 0, 1 \rangle_1$ &
$\big( | 0, 0; 0, 1 \rangle_1 + \sqrt{3}| 1, 1; 0, 1 \rangle_1 \big)/2$ &
$M_{00} - 3\kappa_{qQ} + B_Q$
\\
$Y_2$ & $1^{--}$ & 
$\big( | 1, 0; 1, 1 \rangle_1 + | 0, 1; 1, 1 \rangle_1 \big)/\sqrt{2}$ &
$| 1, 1; 1, L' \rangle_1$ &
$M_{00} - \kappa_{qQ} + 2a + B_Q$
\\
$Y_3$ & $1^{--}$ & 
$| 1, 1; 0, 1 \rangle_1$ &
$\big( \sqrt{3}| 0, 0; 0, 1 \rangle_1 -| 1, 1; 0, 1 \rangle_1 \big)/2$ &
$M_{00} + \kappa_{qQ} + B_Q$
\\
$Y_4$ & $1^{--}$ & 
$| 1, 1; 2, 1 \rangle_1$ &
$| 1, 1; 2, L' \rangle_1$ &
$M_{00} + \kappa_{qQ} + 6a + B_Q$
\\
$Y_5$ & $1^{--}$ & 
$| 1, 1; 2, 3 \rangle_1$ &
$| 1, 1; 2, L' \rangle_1$ &
$M_{00} + \kappa_{qQ} + 16a + 6B_Q$
\\
\botrule
\end{tabular}
}
\label{ali:tbl2}
\end{center}
\end{table}
\vspace*{-3mm}

\noindent The parameters appearing on the r.h. columns of Tables~\ref{ali:tbl1} and~\ref{ali:tbl2} can be
determined using the masses of some of the observed $X,Y,Z$ states, and their
numerical values are given in Table~\ref{ali:tbl3}. Some parameters in
the $c\bar{c}$ and $b\bar{b}$ sectors can also be related using the heavy quark mass
scaling~\cite{Ali:2014dva}. 

\begin{table}
\begin{center}
\tbl{Numerical values of the parameters in $H_{\rm eff}$.}
{\begin{tabular}{c|cc}
\toprule
& charmonium-like & bottomonium-like
\\
\colrule
$M_{00}$ [MeV] & 3957 & 10630
\\
$\kappa_{qQ}$ [MeV] & 67 & 23
\\
$B_Q$ [MeV] & 268 & 329
\\
$a$ [MeV] & 52.5 & 26
\\
\botrule
\end{tabular}
}
\label{ali:tbl3}
\end{center}
\end{table}

\vspace*{-3mm}
\begin{table}
\begin{center}
\tbl{$X,Y,Z$ hadron masses from experiments and in the diquark-model.}
{\begin{tabular}{c|c|cc|cc}
\toprule
&&
\multicolumn{2}{c|}{charmonium-like} & 
\multicolumn{2}{c}{bottomonium-like}
\\
Label & $J^{PC}$ & State & Mass [MeV] & State & Mass [MeV] 
\\
\hline
$X_0$ & $0^{++}$ & 
--- & 3756 &
--- & 10562
\\
$X_0'$ & $0^{++}$ &
--- & 4024 &
--- & 10652
\\
$X_1$ & $1^{++}$ & 
$X(3872)$ & 3890 &
--- & 10607
\\
$Z$ & $1^{+-}$ & 
$Z_c^+(3900)$ & 3890 &
$Z_b^{+,0}(10610)$ & 10607
\\
$Z'$ & $1^{+-}$ &
$Z_c^+(4020)$ & 4024 &
$Z_b^+(10650)$ & 10652
\\
$X_2$ & $2^{++}$ &
--- & 4024 &
--- & 10652
\\
\hline
$Y_1$ & $1^{--}$ &
$Y(4008)$ & 4024 &
$Y_b(10891)$ & 10891
\\
$Y_2$ & $1^{--}$ &
$Y(4260)$ & 4263 &
$Y_b(10987)$ & {\bf 10987}
\\
$Y_3$ & $1^{--}$ &
$Y(4290)$ (or $Y(4220)$) & 4292 &
--- & {\bf 10981}
\\
$Y_4$ & $1^{--}$ &
$Y(4630)$ & 4607 &
--- & 11135
\\
$Y_5$ & $1^{--}$ &
--- & 6472 &
--- & 13036
\\
\botrule
\end{tabular}
}
\label{ali:tbl4}
\end{center}
\end{table}
\vspace*{3mm}
\noindent Typical errors on the masses due to parametric uncertainties are estimated to be about 30 MeV.
As we see from table~\ref{ali:tbl4}, there are lot more $X,Y,Z$ hadrons observed in experiments
in the charmonium-like sector than in the bottomonium-like sector, with essentially three entries
$Z_b^+(10610)$, $Z_b^+(10650)$ and $Y_b(10891)$ in the latter case. There are several predictions in
the charmonium-like sector, which, with the values of the parameters given in the tables above, are in the
right ball-park~\footnote{I thank Satoshi Mishima for providing these estimates.}.
 It should be remarked that these input values, in particular for the quark-quark
couplings in a diquark, $\kappa_{qQ}$,  are larger than in the earlier determinations of
the same by Maiani {\it et al.}~\cite{Maiani:2004vq}. Better agreement is reached with experiments assuming that
diquarks are more tightly bound than suggested from the analysis of the baryons
 in the diquark-quark picture, and the spectrum shown here is in agreement with the
one in the modified scheme\cite{Maiani:2014aja}. Alternative calculations of the tetraquark spectrum
based on diquark-antidiquark model have been carried out~\cite{Ebert:2005nc}.

The exotic bottomonium-like states are currently rather sparse.
The reason for this is that quite a few exotic charmonium-like states were observed in the decays of 
$B$-hadrons. This mode is obviously not available for the hidden $b\bar{b}$ states. They can only be produced
in hadro- and electroweak  high energy processes. Tetraquark states
with a single $b$ quark can, in principle, also be produced in the decays of the $B_c$ mesons, as pointed out
 recently~\cite{Ali:2016gdg}.
As the $c\bar{c}$ and $b\bar{b}$ cross-section at the LHC are very large, we 
anticipate that the exotic spectroscopy involving the open  and hidden heavy quarks is an area
where significant new results  will be reported by all the LHC experiments. Measurements of the
production and decays of exotica, such as transverse-momentum distributions and polarization information,
will go a long way in understanding the underlying dynamics. 

As a side remark, we mention that recently there has been a lot of excitement due to the  
D0 observation~\cite{D0:2016mwd} of a narrow structure $X(5568)$, consisting of four different quark flavors
 ${\tt {\it (b d u s)}}$,
found through the $B_s^0 \pi^\pm$ decay mode. However, this has not been confirmed by the LHCb 
collaboration~\cite{LHCb:2016ppf},
despite the fact that LHCb has 20 times higher $B_s^0$ sample than that of D0. This would have been the
first discovery of an open $b$-quark tetraquark state. They are anticipated in the compact tetraquark
picture~\cite{Ali:2016gdg}, and also in the hadron molecule framework~\cite{Agaev:2016urs}. We wait for more
data from the LHC experiments.
 
We now discuss the three observed exotic states in the bottomonium sector in detail.
The hidden $b \bar{b}$ state
$Y_b(10890)$ with  $J^{\rm P}=1^{--}$  was discovered by Belle in 2007~\cite{Abe:2007tk} in the
 process $e^+e^- \to Y_b(10890) \to (\Upsilon(1S), \Upsilon(2S), \Upsilon(3S)) \pi^+ \pi^-$ just above the
$\Upsilon(5S)$. The branching ratios measured are about two orders of magnitude larger than anticipated
from similar dipionic transitions in the lower $\Upsilon(nS)$ states and $\psi^\prime$
(for a review and references to earlier work, see Brambilla {\it et al}~\cite{Brambilla:2010cs}.).
Also the dipion invariant mass distributions in the decays of $Y_b$ are  marked by the presence of
the resonances $f_0(980)$ and 
$f_2(1270)$. This state
was interpreted as a $J^{\rm PC}=1^{--}$ P-wave tetraquark~\cite{Ali:2009pi,Ali:2009es}.
Subsequent to this, a Van Royen-Weiskopf formalism was used~\cite{Ali:2010pq}
 in which direct electromagnetic couplings with
the diquark-antidiquark pair of the $Y_b$ was assumed. Due to the $P$-wave nature of the
$Y_b(10890)$, with a commensurate  small overlap function,  the observed  small
production cross-section in $e^+e^- \to b\bar{b}$ was explained. In the tetraquark picture,
$Y_b(10890)$ is the $b\bar{b}$ analogue of the $c\bar{c}$ state $Y_c(4260)$, also a $P$-wave, which  
is likewise found to have  a very small production cross-section, but decays readily into $J/\psi \pi^+\pi^-$.
Hence, the two have very similar production and decay characteristics, and, in all likelihood, they have 
similar compositions.

The current status of $Y_b(10890)$ is unclear. Subsequent to the discovery of $Y_b(10890)$, Belle undertook
high-statistics scans for the ratio
$R_{b\bar{b}}=\sigma(e^+e^- \to b\bar{b})/\sigma (e^+ e^- \to \mu^+ \mu^-)$, and also measured more precisely the
ratios $R_{\Upsilon(nS) \pi^+\pi^-}$. No results are available on $R_{\Upsilon(nS) \pi^+\pi^-}$ from BaBar, so we discuss
the analysis reported by Belle.
 The two masses, $M(5S)_{b\bar{b}}$ measured through $R_{b\bar{b}}$, and $M(Y_b)$,
 measured through $R_{\Upsilon(nS) \pi^+\pi^-}$, now differ by slightly more than 2$\sigma$, yielding 
$M(5S)_{b\bar{b}} -M(Y_b)= -9 \pm 4$ MeV.
 From the mass difference alone, these two could very well be just one and the same state,
namely the canonical $\Upsilon(5S)$ - an interpretation adopted by the Belle
 collaboration~\cite{Santel:2015qga}. On the other hand, it is now the book keeping of the branching ratios
measured at or near the $\Upsilon(5S)$, which is puzzling. This is  reflected in the paradox that
{\it direct production} of the $B^{(*)}\bar{B}^{(*)}$ as well as of $B_s\bar{B}_s^{(*)}$ states have
essentially  no place
in the Belle counting~\cite{Santel:2015qga}, as the branching
ratios of the $\Upsilon(5S)$ are already saturated by the exotic states
 $(\Upsilon(nS) \pi^+\pi^-, h_b(mP) \pi^+\pi^-,
Z_b(10610)^\pm \pi^\mp, Z_b(10650)^\pm \pi^\mp$ and their isospin partners). In our opinion, an interpretation
of the Belle data based on two resonances $\Upsilon(5S)$ and $Y_b(10890)$ is more natural, with
$\Upsilon(5S)$ having the decays expected for the bottomonium $S$-state above the $B^{(*)}\bar{B}^{(*)}$ threshold,
and the decays of $Y_b(10890)$, a tetraquark, being the source of the exotic states seen.     
 As  data taking starts in a
couple of years in the form of a new and expanded collaboration, Belle-II, cleaning up the current analysis in 
the $\Upsilon(5S)$ and $\Upsilon(6S)$ region should be one of their top priorities. In the meanwhile, the 2007
discovery of $Y_b(10890)$ stands, not having been retracted by Belle, at least as far as I know.

\begin{figure}
\centerline{\includegraphics[width=12cm]{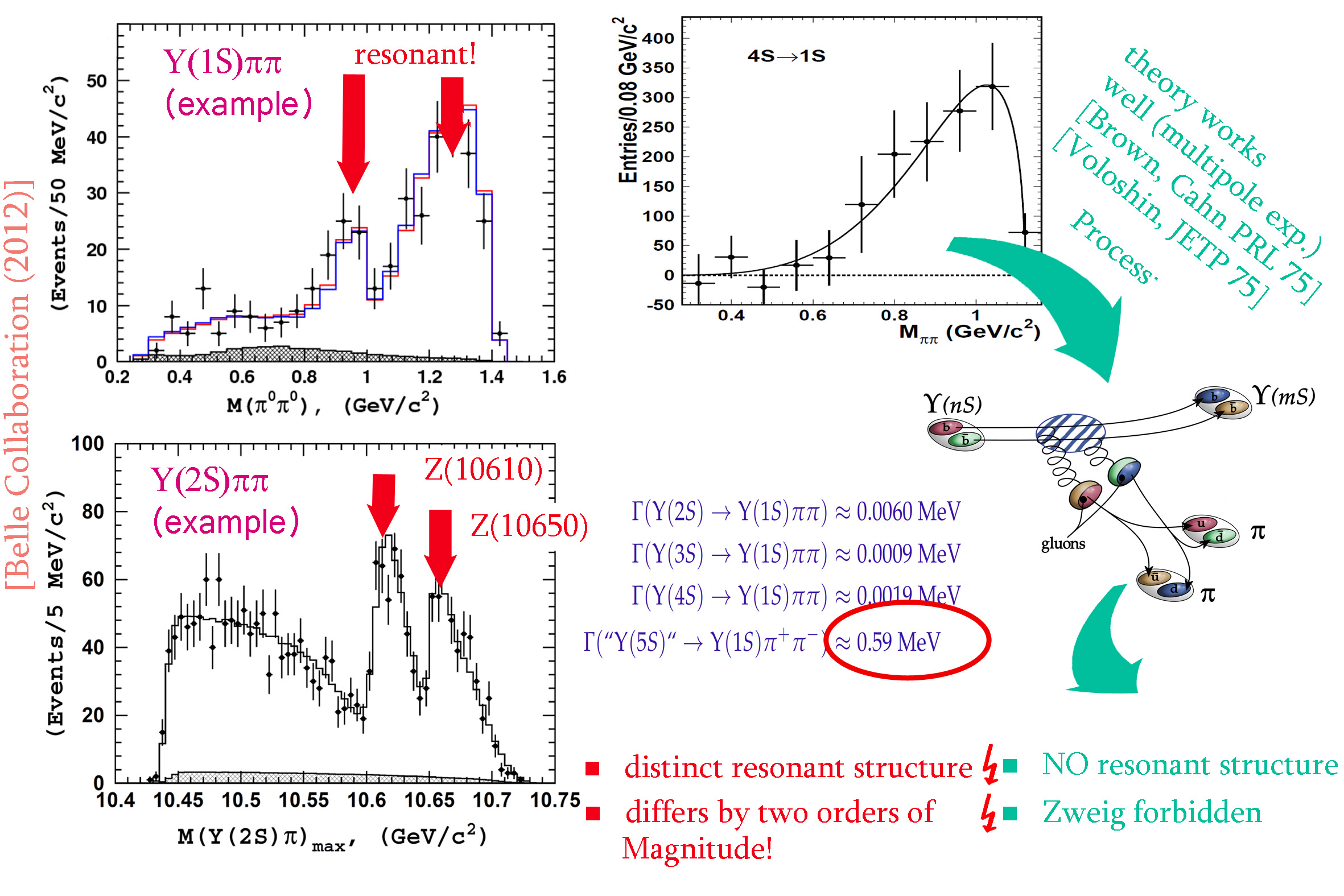}}
\vspace{-3mm}
\caption{Dipion invariant mass distribution in $\Upsilon(10890) \to \Upsilon(1S) \pi^0 \pi^0$ (upper left frame);
 the resonances indicated
 in the dipion spectrum correspond to the $f_0(980)$ and $f_2(1270)$;
the resonances $Z(10610)$ and $Z(10650)$ are indicated in
the $\Upsilon(2S) \pi^+$ invariant mass distribution from $\Upsilon(10890) \to \Upsilon(2S) \pi^+ \pi^-$ (lower left frame).
 The data are from the Belle collaboration~\cite{Belle:2011aa}.
The upper right hand frame shows the dipion invariant mass distribution in $\Upsilon(4S) \to \Upsilon(1S) \pi^+\pi^-$,
and the theoretical curve (with the references) is based on  the Zweig-forbidden process shown below. The measured decay widths 
from $\Upsilon(nS) \to \Upsilon(1S)\pi^+ \pi^-$ $nS=2S,3S,4S$ and $\Upsilon(10890) \to \Upsilon(1S) \pi^+\pi^-$
are also shown.This figure serves to underscore the drastically different underlying mechanisms for dipionic
transitions in $\Upsilon(nS)$ and $\Upsilon(10890)$ decays.}
\label{ali:fig4}
\end{figure}

 Thus, there is a good case that 
 $\Upsilon(5S)$ and $Y_b(10890)$, while having the same $J^{\rm PC}=1^{--}$ quantum numbers and almost the same mass,
are {\it different} states. As already mesntioned, this is hinted by the
drastically different decay characteristics of the dipionic transitions involving the
 lower quarkonia $S$-states, such as $\Upsilon(4S) \to \Upsilon(1S) \pi^+\pi^-$, on one hand, and similar decays of the
$Y_b$, on the other. These anomalies are seen both in the decay rates and in the dipion invariant mas spectra in the
$\Upsilon(nS)\pi^+\pi^-$ modes. The large branching ratios of $Y_b \to \Upsilon(nS) \pi^+\pi^-$,
 as well as of $Y(4260) \to J/\psi \pi^+\pi^-$, are due to the Zweig-allowed nature of these transitions,
as the initial and final states have the same valence quarks. The final state $\Upsilon(nS) \pi^+\pi^-$ in
$Y_b$ decays requires the excitation of a $q\bar{q}$ pair from the vacuum. Since, the light scalars
$\sigma_0$, $f_0(980)$ are themselves tetraquark candidates~\cite{Hooft:2008we,Fariborz:2008bd}, 
 they are expected to show up
 in the $\pi^+\pi^-$
invariant mass distributions, as opposed to the corresponding spectrum in the transition
$\Upsilon(4S) \to \Upsilon(1S) \pi^+\pi^-$ (see Fig.~\ref{ali:fig4}). 
 Subsequent discoveries~\cite{Belle:2011aa} of the charged states $Z_b^+(10610)$ and
$Z_b^+(10650)$, found in the decays $\Upsilon(5s)/Y_b \to Z_b^+(10610) \pi^-, Z_b^+(10650) \pi^-$, 
leading to the final states $\Upsilon(1S) \pi^+\pi^-$, $\Upsilon(2S) \pi^+\pi^-$, $\Upsilon(3S) \pi^+\pi^-$,
$h_b(1P)\pi^+\pi^-$ and $h_b(2P) \pi^+\pi^-$, give credence to the tetraquark interpretation, as discussed below.

\subsection{ Heavy-Quark-Spin Flip in $\Upsilon(10890) \to h_b(1P, 2P) \pi \pi$}%
We summarize the relative rates and strong phases measured by Belle~\cite{Belle:2011aa} in the process $\Upsilon(10890) \to
 \Upsilon(nS) \pi^+\pi^-, h_b(mP) \pi^+\pi^-$, with $n=1,2,3$ and $m=1,2$ in Table~\ref{ali:tbl5}.
For ease of writing we shall use the notation $Z_b$ and $Z_b^\prime$ for the two charged $Z_b$ states. Here no assumption is
made about the nature of $\Upsilon(10890)$, it can be either $\Upsilon(5S)$ or $Y_b$. Of these, the decay $\Upsilon(10890)
\to \Upsilon(1S) \pi^+\pi^-$ involves both a resonant (i.e., via $Z/Z^\prime$) and a direct component, but the other four
are dominated by the resonant contribution. One notices that the relative normalizations are very similar and
the phases of the $(\Upsilon(2S), \Upsilon(3S)) \pi^+\pi^-$ differ by about $180^\circ$ compared to
the ones in $(h_b(1P, h_b(2P))\pi^+\pi^-$. At the first sight this seems to violate the heavy-quark-spin conservation,
as in the initial state $s_{b\bar{b}}=1$, which remains unchanged for the $\Upsilon(nS)$ in the final state, i.e., it
involves an $s_{b\bar{b}}=1 \to s_{b\bar{b}}=1$ transition, but as  $s_{b\bar{b}}=0$ for the $h_b(mP)$, this involves
an $s_{b\bar{b}}=1 \to s_{b\bar{b}}=0$ transition, which should have been suppressed, but is not
 supported by data. 

\begin{table}
\begin{center}
\tbl{Relative normalizations and  phases for
 $s_{b\bar{b}}: 1 \to 1$~ and $1 \to 0$~transitions in 
$\Upsilon(10890)$ decays~\cite{Belle:2011aa}.}
{\begin{tabular}{| l | c | c | c | c | c | }
\toprule \\[-4.5mm]
    & & & & &  \\[-2mm]
   \hspace{-2mm} Final State \hspace{-2mm} & 
$\Upsilon(1S)\pi^+\pi^-$ & $\Upsilon(2S)\pi^+\pi^-$ & $\Upsilon(3S)\pi^+\pi^-$ &  
$h_b(1P)\pi^+\pi^-$ & \hspace{-2mm} $h_b(2P)\pi^+\pi^-$ \hspace{-2mm} \\
    & & & & &  \\
   \hline\hline
    & & & & &  \\
 \hspace{-2mm} Rel. Norm.\hspace{-2mm} & 
\hspace{-2mm} $0.57\pm 0.21^{+0.19}_{-0.04}$ \hspace{-2mm} & 
\hspace{-2mm} $0.86\pm 0.11^{+0.04}_{-0.10}$ \hspace{-2mm} & 
\hspace{-2mm} $0.96\pm 0.14^{+0.08}_{-0.05}$ \hspace{-2mm} & 
\hspace{-2mm} $1.39\pm 0.37^{+0.05}_{-0.15}$ \hspace{-2mm} & 
\hspace{-4mm} $1.6^{+0.6+0.4}_{-0.4-0.6} $ \hspace{-4mm} \\
   & & & & &  \\
  \hline
    & & & & & \\
  \hspace{-2mm} Rel. Phase \hspace{-3mm} & 
$58\pm43^{+4}_{-9}$ & $-13\pm13^{+17}_{-8}$ & $-9\pm 19^{+11}_{-26}$ &  $187^{+44+3}_{-57-12}$ & 
\hspace{-4mm} $181^{+65+74}_{-105-109}$ \hspace{-4mm} \\[-1mm]
 & & & & &  \\[-1mm]
\botrule 
\end{tabular}
}
\label{ali:tbl5}
\end{center}
\end{table}

\vspace{-1mm}
\noindent It has been shown that this contradiction is only apparent~\cite{Ali:2014dva}.
 Expressing the states $Z_b$~and $Z_b^\prime$~in the
basis of definite $b\bar{b}$~and light quark $q\bar{q}$~spins, it becomes
evident that both the $Z_b$ and $Z_b^\prime$ have $s_{b\bar{b}}=1$ and $s_{b\bar{b}}=0$
components, 
\begin{eqnarray}
&&|Z_b\rangle=\frac{|1_{q\bar q},0_{b\bar b}\rangle-|0_{q\bar q},1_{b\bar b}\rangle}{\sqrt{2}},
~~|Z_b^\prime\rangle=\frac{|1_{q\bar q},0_{b\bar b}\rangle+|0_{q\bar q},1_{b\bar b}\rangle}{\sqrt{2}}\nonumber.
\end{eqnarray}
 Defining ($g$~is the effective couplings at the vertices $\Upsilon\, Z_b\, \pi$ and $Z_b\, h_b\, \pi$)
\begin{eqnarray}
&&g_{Z}\equiv g(\Upsilon \to Z_b\pi)g(Z_b\to h_b\pi)\propto -\alpha\beta\langle h_b|Z_b\rangle \langle Z_b|\Upsilon\rangle,\nonumber\\
&&g_{Z^\prime}\equiv g(\Upsilon \to Z_b^\prime\pi)g(Z_b^\prime \to h_b\pi)\propto \alpha\beta\langle h_b|Z^\prime_b\rangle
 \langle Z^\prime_b|\Upsilon\rangle\nonumber,\end{eqnarray}
we note that within errors, Belle data is consistent with the heavy quark spin conservation,
which requires~$g_Z=-g_{Z^\prime}$. The two-component nature of the $Z_b$ and $Z_b^\prime$ is also the feature
which was pointed out earlier for $Y_b$ in the context of the direct transition $Y_b(10890) \to \Upsilon(1S) \pi^+\pi^-$.
%
 To determine the coefficients $\alpha$~and $\beta$, one has to resort to 
$s_{b\bar{b}}$: $1 \to 1 $~transitions
%
\begin{equation}
 \Upsilon(10890)\to  Z_b/Z_b^\prime+\pi\to  \Upsilon(nS)\pi\pi~(n=1,2,3).\nonumber
 \end{equation}
The analogous effective couplings are
\begin{eqnarray}
&&f_{Z}=f(\Upsilon \to Z_b\pi)f(Z_b\to \Upsilon(nS)\pi)\propto |\beta|^2 \langle \Upsilon(nS)|0_{q\bar q},1_{b\bar b}\rangle \langle 0_{q\bar q},1_{b\bar b}|\Upsilon\rangle,\notag \\ 
&&f_{Z^\prime}=f(\Upsilon \to Z_b^\prime\pi)f(Z_b^\prime\to \Upsilon(nS)\pi)\propto |\alpha|^2 \langle \Upsilon(nS)|0_{q\bar q},1_{b\bar b}\rangle \langle 0_{q\bar q},1_{b\bar b} |\Upsilon\rangle.\notag 
\end{eqnarray}
 Dalitz analysis indicates that
 $\Upsilon(10890)\to  Z_b/Z_b^\prime+\pi\to  \Upsilon(nS)\pi\pi~(n=1,2,3)$~proceed
mainly through the resonances $Z_b$~and $Z_b^\prime$, though
 $\Upsilon(10890) \to  \Upsilon(1S)\pi\pi$~has a significant direct component,
expected in tetraquark interpretation of $\Upsilon(10890)$~\cite{Ali:2010pq}.
A comprehensive analysis of the Belle data including the direct and resonant
components is required to test the underlying dynamics, which yet to be carried out.
However,  parametrizing the amplitudes in terms of two Breit-Wigners, one can determine the
ratio $\alpha/\beta$ from 
$\Upsilon(10890)\to  Z_b/Z_b^\prime+\pi\to  \Upsilon(nS)\pi\pi~(n=1,2,3)$. For the $s_{b\bar b}:1\to 1~{\rm transition}$, we get
for the averaged quantities:
\begin{eqnarray}
\overline{{\rm Rel. Norm.}}= 0.85\pm 0.08=|\alpha|^2/|\beta|^2;
~~\overline{{\rm Rel. Phase}}= (-8\pm10)^\circ.\nonumber
\end{eqnarray}
For the $s_{b\bar b}:1\to 0~{\rm transition}$, we get
\begin{eqnarray}
\overline{{\rm Rel. Norm.}}= 1.4\pm 0.3;
~~\overline{{\rm Rel. Phase}}= (185 \pm 42)^\circ.\nonumber
\end{eqnarray}
Within errors, the tetraquark assignment  with $\alpha=\beta=1$~is
 supported, i.e.,
\begin{eqnarray}
&&|Z_b\rangle=\frac{|1_{bq},0_{\bar b\bar q}\rangle-|0_{bq},1_{\bar b\bar q}\rangle}{\sqrt{2}},
~~|Z_b^\prime\rangle=|1_{b q},1_{\bar b\bar q}\rangle_{J=1},\nonumber
\end{eqnarray}
and
\begin{eqnarray}
&&|Z_b\rangle=\frac{|1_{q\bar q},0_{b\bar b}\rangle-|0_{q\bar q},1_{b\bar b}\rangle}{\sqrt{2}},
~~|Z_b^\prime\rangle=\frac{|1_{q\bar q},0_{b\bar b}\rangle+|0_{q\bar q},1_{b\bar b}\rangle}{\sqrt{2}}.\nonumber
\end{eqnarray}
It is interesting that similar conclusion was drawn in the `molecular' interpretation~\cite{Bondar:2011ev}
 of the $Z_b$ and $Z_b^\prime$.

\noindent The Fierz rearrangement used in obtaining second of the above relations would put together the $b\bar{q}$ and $q\bar{b}$ fields,
yielding 

\begin{eqnarray}
&&|Z_b\rangle=|1_{b \bar{q}},1_{\bar {b} q}\rangle_{J=1},
~~|Z_b^\prime \rangle=\frac{|1_{b\bar{q}},0_{q \bar{b}}\rangle+|0_{b\bar{q}},1_{q \bar{b}}\rangle}{\sqrt{2}}\nonumber.
\end{eqnarray}
\noindent Here, the labels $0_{b\bar{q}}$ and $1_{\bar{q}b}$ could be viewed as indicating $B$ and $B^*$ mesons, respectively,
leading to the prediction $Z_b \to B^* \bar{B}^*$ and $Z_b^\prime \to B \bar{B}^*$, which is not in agreement
 with the Belle data~\cite{Belle:2011aa}.
However, this argument rests on the conservation of the light quark spin, for which there is no theoretical
 foundation. Hence,
this last relation is not reliable. Since $Y_b(10890)$ and $\Upsilon(5S)$ are rather close in mass, and there
 is an issue with
the unaccounted {\it direct production} of the $B^* \bar{B}^*$ and $B \bar{B}^*$ states in the Belle data collected 
in their vicinity, we conclude that the experimental situation is still in a state of flux and
look forward to its resolution with the consolidated  Belle-II data.

\subsection{Drell-Yan mechanism for vector exotica production at the LHC and Tevatron}
The exotic hadrons having $J^{\rm PC}=1^{--}$ can be produced at the Tevatron and LHC via the Drell-Yan
 process~\cite{Ali:2011qi}
$pp (\bar{p}) \to \gamma^* \to V +...$. The cases $V=\phi(2170), Y(4260), Y_b(10890)$ have been studied.
  With the other two hadrons already discussed earlier, we recall that 
 $\phi(2170)$ was first observed in the ISR process $e^+e^- \to \gamma_{\rm ISR} f_0(980) \phi(1020)$ by
 BaBaR~\cite{Aubert:2006bu} and later confirmed by BESII~\cite{Ablikim:2007ab} and Belle~\cite{Shen:2009zze}.
 Drenska {\it et al.}~\cite{Drenska:2008gr} interpreted  $\phi(1270)$ as a P-wave tetraquark
 $[sq][\bar{s} \bar{q}]$. Thus,  all three vector exotica are assumed to be the first orbital 
excitation of diquark-antidiquark states with a hidden $s\bar{s}$, $c\bar{c}$ and $b\bar{b}$ quark content,
respectively. As all three have very small branching ratios in a dilepton pair, they should be searched
for in the decay modes in which they have been discovered, and these are $\phi(2170) \to f_0(980)\phi(1020)
\to \pi^+\pi^- K^+ K^-$, $Y(4260) \to J/\psi \pi^+\pi^- \to \mu^+\mu^- \pi^+ \pi^-$ and
$Y_b(10890) \to \Upsilon(nS) \pi^+\pi^- \to \mu^+\mu^- \pi^+\pi^-$. Thus, they involve four charged particles,
which can be detected at hadron colliders. With their masses, total and partial decay widths taken from the
PDG~\cite{Olive:2014pdg}, the cross sections  for the processes
$p \bar{p}(p) \to \phi(2170) (\to  \phi(1020) f_0(980) \to K^+K^- \pi^+\pi^-)$,
  $p \bar{p}(p) \to Y(4260)(\to  J/\psi \pi^+\pi^- \to \mu^+\mu^- \pi^+\pi^-)$, and
 $ p \bar{p}(p) \to  Y_b(10890) (\to  \Upsilon (1S,2S,3S)\pi^+\pi^- \to \mu^+\mu^- \pi^+\pi^-)$,  
 at the Tevatron ($\sqrt s=$ 1.96 TeV) and the LHC are given in Table~\ref{ali:tbl6}, with the indicated
rapidity ranges. All these processes have measurable rates, and they
should be searched for, in particular, at the LHC. 
\begin{table}
\begin{center}
\tbl{Cross sections  (in units of pb)  for the processes 
$p \bar{p}(p) \to \phi(2170) (\to  \phi(1020) f_0(980) \to K^+K^- \pi^+\pi^-)$,
  $ p \bar{p}(p) \to Y(4260)(\to  J/\psi \pi^+\pi^- \to \mu^+\mu^- \pi^+\pi^-)$, and
 $ p \bar{p}(p) \to  Y_b(10890) (\to  \Upsilon (1S,2S,3S)\pi^+\pi^- \to \mu^+\mu^- \pi^+\pi^-)$,  
  at the Tevatron ($\sqrt s=$ 1.96 TeV) and the LHC~\cite{Ali:2011qi}.\hspace{29mm} }
{\begin{tabular}{|c|c|c|c|} 
\toprule \\[-3mm]
&$\phi(2170)$ &$ Y(4260)$ 
&$Y_{b}(10890)$  \\[-3mm]
\\\colrule 
 Tevatron$(|y|<2.5)$  &$2.3^{+0.9}_{-0.9}$ &$0.23^{+0.19}_{-0.05}$    &$0.0020^{+0.0006}_{-0.0005} $
\\[2mm]
LHC 7TeV $(|y|<2.5)$  &$3.6^{+1.4}_{-1.4}$ &$0.40^{+0.32}_{-0.09}$  &$0.0040^{+0.0013}_{-0.0011}$ 
\\[2mm] 
LHCb 7TeV   ($1.9<y<4.9$) &$2.2^{+1.2}_{-1.1}$&$0.24^{+0.20}_{-0.07}$ &$0.0023^{+0.0007}_{-0.0006}$ 
\\[2mm] 
LHC 14TeV $(|y|<2.5)$  &$4.5^{+1.9}_{-1.9}$ &$0.54^{+0.44}_{-0.12}$  &$0.0060^{+0.0019}_{-0.0016}$ 
\\[2mm]
LHCb  14TeV  ($1.9<y<4.9$) &$2.7^{+1.9}_{-1.6}$&$0.31^{+0.27}_{-0.11}$ &$0.0033^{+0.0011}_{-0.0010}$ 
\\
\botrule
\end{tabular} 
}
\label{ali:tbl6}
\end{center}
\end{table}

Summarizing this discussion, we note that there are several puzzles in the $X,Y,Z$ sector.
These involve the nature of the
$J^{PC}=1^{--}$ states, $Y(4260)$ and  $Y(10890)$, and whether they are related with each
other. Also, whether $Y(10890)$ and $\Upsilon(5S)$ are one and the same particle is still an open issue. 
 In principle, both  $Y(4260)$ and  $Y(10890)$ can be produced at the LHC and measured through
 the $J\psi \pi^+\pi^-$ and
$\Upsilon(nS) \pi^+\pi^-$ $(nS=1S,2S,3S)$ modes, respectively. Their hadroproduction cross-sections are unfortunately
 uncertain, but their (normalized) transverse momentum distributions will be quite revealing.
 As they are both $J^{PC}=1^{--}$ hadrons, they
can also be produced via the Drell-Yan mechanism and detected through their signature decay modes.
 We have argued that the tetraquark interpretation of the charged
exotics $Z_b$ and $Z_b^\prime$ leads to a straight forward understanding of the relative rates
and strong phases of the heavy quark
 spin non-flip and
 spin-flip transitions in the decays $\Upsilon(10890) \to \Upsilon(nS) \pi^+\pi^-$ and 
 $\Upsilon(10890) \to h_b(mP) \pi^+\pi^-$, respectively. In the tetraquark picture, 
the corresponding hadrons in the charm sector $Z_c$ and $Z_c^\prime$ are related to their $b\bar{b}$ counterparts.
We look forward to the higher luminosity data at Bell-II and LHC to resolve some of these issues.

\section{Pentaquarks}
Pentaquarks remained cursed for almost a decade under the shadow of the botched discoveries
 of $\Theta(1540),\,\, \Phi(1860),\,\,\Theta_c(3100)$. The sentiment of the particle physics
community is reflected in the terse 2014 PDG review~\cite{PDG-Wohl}:\\
{\it There are two or three recent experiments that find weak evidence for signals near the nominal masses,
 but there is simply no point in tabulating them in view of the overwhelming evidence that the claimed
 pentaquarks do not exist. The only advance in particle physics thought worthy of mention in 
the American Institute of Physics ``Physics News in 2003'' was a false alarm. The whole story --- is 
a curious episode in the history of science.}

This seems to have changed by the observation of $J/\psi p$ resonances consistent with pentaquark states in
$\Lambda_b^0 \to J/\psi K^- p$ decays by the LHCb collaboration~\cite{Aaij:2015tga}.
 The discovery channel ($\sqrt{s}=7$ and 8 TeV,
$\int Ldt= 3$ fb$^{-1}$) is
\vspace*{-3mm}
\begin{eqnarray}
&& pp \to b\bar{b} \to \Lambda_b X; \Lambda_b \to K^- J/\psi p. \nonumber
\end{eqnarray}
A statistically good fit of the $m_{J/\psi p}$-distribution is consistent with the presence of two resonant
states, henceforth called $P_c(4450)^+$ and $P_c(4380)^+$, with the following characteristics
\vspace*{-2mm}
\begin{eqnarray}
&&  M=4449.8 \pm 1.7 \pm 2.5~{\rm MeV};~~\Gamma=39 \pm 5 \pm 19~{\rm MeV}, \nonumber
\end{eqnarray}
and
\vspace*{-3mm}
\begin{eqnarray}
&&  M=4380 \pm 8 \pm 29~{\rm MeV};~~\Gamma=205 \pm 18 \pm 86~{\rm MeV}, \nonumber
\end{eqnarray}
having the statistical significance of 12$\sigma$ and 9$\sigma$, respectively.
 Both of them  carry a unit of baryonic number and have the valence quarks
$ P_c^+ = \bar c c u u d$. The preferred $J^{\rm P}$ 
assignments are $5/2^+$ for the $P_c(4450)^+$ and $3/2^-$ for the $P_c(4380)^+$. Doing an Argand-diagram analysis in
the (${\rm Im}~A^{P_c}$ - ${\rm Re}~A^{P_c}$) plane, the phase change in the amplitude is consistent
with a resonance for the $P_c(4450)^+$, but less so for the $P_c(4380)^+$. 

Following a pattern seen for the tetraquark candidates, namely their proximity to respective thresholds,
such as $D\bar{D}^*$ for the $X(3872)$, $B\bar{B}^*$ and $B^*\bar{B}^*$ for the $Z_b(10610)$ and
$Z_b(10650)$, respectively, also the two pentaquark candidates $P_c(4380)$ and $P_c(4450)$ lie close to several charm
meson-baryon thresholds~\cite{Burns:2015dwa}. The $\Sigma_c^{*+} \bar{D}^0$ has a threshold
 of $4382.3 \pm 2.4$ MeV, tantalizingly
close to the mass of $P_c(4380)^+$. In the case of $P_c(4450)^+$, there are several thresholds within
striking distance, $\chi_{c1}p (4448.93 \pm 0.07), \Lambda_c^{*+} \bar{D}^0 (4457.09 \pm 0.35),
\Sigma_c^+ \bar{D}^{*0} (4459.9 \pm 0.9)$, and $\Sigma_c^+ \bar{D}^0 \pi^0 (4452.7 \pm 0.5)$, where the masses are
in units of MeV. This has led to a number of hypotheses to accommodate the two $P_c$ states,
which can be classified under four different mechanisms:
\vspace*{-3mm}
\begin{itemize}
\item Rescattering-induced kinematic effects~\cite{Guo:2015umn,Liu:2015fea,Mikhasenko:2015vca,Meissner:2015mza}.
\item $P_c(4380)$ and $P_c(4450)$  as baryocharmonia~\cite{Kubarovsky:2015aaa}.
\item Open charm-baryons and charm-meson bound 
states~\cite{Chen:2015moa,He:2015cea,Roca:2015dva,Chen:2015loa,Xiao:2015fia}.
 \item Compact pentaquarks~\cite{Maiani:2015vwa,Lebed:2015tna,Li:2015gta,Mironov:2015ica,Anisovich:2015cia,Ghosh:2015ksa,Wang:2015epa,Wang:2015ava}
\end{itemize}  
\vspace*{-3mm}
We discuss the first three briefly and the compact pentaquarks in somewhat more detail subsequently.

Kinematic effects can result in a narrow
structure around the $\chi_{c1}p$ threshold. Two possible mechanisms shown in Fig.~\ref{ali:fig7} are:\\
 (a) 2-point loop with a 3-body
production~$\Lambda_b^0 \to K^- \,\chi_{c1}\, p $~followed by the rescattering
 process~$\chi_{c1}\, p \to J/\psi\, p$, and\\ 
(b) in which $K^-\,p $~is produced from an intermediate~$\Lambda^*$~and the proton rescatters
 with the~$\chi_{c1}$~into a~{$J/\psi \, p$, as shown below.

\begin{figure}
\centerline{\includegraphics[width=0.95\textwidth]{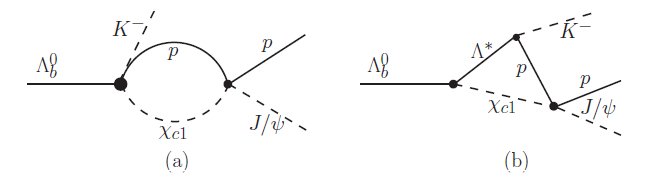}}
\vspace{-2mm}
\caption{The two scattering diagrams discussed in the text~\cite{Guo:2015umn}.}
\label{ali:fig7}
\end{figure}
In the baryocharmonium picture, the $P_c$ states are hadroquarkonium-type composites of $J/\psi$ and excited
nucleon states similar to the known resonances $N(1440)$ and $N(1520)$.  Photoproduction
of the $P_c$ states in $\gamma + p$ collisions is advocated as sensitive probe of this mechanism~\cite{Kubarovsky:2015aaa}.

In the hadronic molecular interpretation, one identifies $P_c(4380)^+$ 
with $\Sigma_c(2455)\bar{D}^*$~and $P_c(4450)^+$~with~$\Sigma_c(2520)\bar{D}^*$,
which are bound by a pion exchange. This can be expressed in terms of the 
effective Lagrangians:
\vspace*{-1mm}
\begin{align}
{\cal L}_{\cal P}  &=ig {\rm Tr} \left[\bar{H}_a^{(\bar{Q})}\, \gamma^\mu \, A^\mu_{ab} \, \gamma_5 H_b^{(\bar{Q})}\right],   \nonumber \\
{\cal L}_{\cal S}  &= -\frac{3}{2} g_1 \epsilon^{\mu \lambda \nu \kappa} v_\kappa {\rm Tr }
\left[\bar{\cal S}_\mu \, A_\nu \, {\cal S}_\lambda \right].   \nonumber
\end{align}
Here $H_a^{(\bar{Q})} = [P_a^{*(\bar{Q})\,\mu}\gamma_\mu - P_a^{(\bar{Q})} \gamma_5](1-\slashed{v})/2$
is a pseudoscalar and  vector charmed meson multiplet~$(D,D^*)$, $v$ being the four-velocity vector
 $v=(0,\vec{1})$, 
 ${\cal S}_\mu= 1/\sqrt{3}(\gamma_\mu + v_\mu) \gamma^5 {\cal B}_6 + {\cal B}^*_{6\mu}$ stands
for the charmed baryon multiplet, with~${\cal B}_6 $ and ${\cal B}^*_{6\mu} $ corresponding to
 the $J^P=1/2^+ $ and $J^P=3/2^+$ in~$6_F$ flavor representation, respectively, and 
 $A_\mu$~is an axial-vector current, containing a pion chiral multiplet.  This interaction lagrangian is
used to work out effective potentials, energy levels and wave-functions
of the $ \Sigma_c^{(*)} \bar{D}^*$ systems. In this picture,
$P_c(4380)^+$ is a $\Sigma_c \bar{D}^*$~$(I=1/2, \, J=3/2)$~molecule, and
$P_c(4450)^+$ is a $\Sigma^*_c \bar{D}^*$~$(I=1/2, \, J=5/2)$ molecule. Apart from accommodating the two
observed pentaquarks, this framework
predicts two additional hidden-charm molecular pentaquark states,
$\Sigma_c \bar{D}^*$~$(I=3/2,\, J=1/2)$~and~$\Sigma^*_c \bar{D}^*$~$(I=3/2,\, J=1/2)$, 
 which are isospin partners of $P_c(4380)^+$~and~$P_c(4450)^+$, respectively, 
decaying into $\Delta(1232) J/\psi$~and~$\Delta(1232) \eta_c$. In addition, 
a rich pentaquark spectrum of states for the hidden-bottom~$ (\Sigma_b B^*, \Sigma^*_b B^* )$~,
$B_c$-like~$ (\Sigma_c B^*, \Sigma^*_c B^*)$~and ~$(\Sigma_b \bar{D}^*, \Sigma_b^* \bar{D}^*)$
with well-defined~{$(I,J)$~are predicted.

\subsection{$P_c(4380)^+$ and $P_c(4450)^+$ as compact pentaquarks}
 This hypothesis  has been put forward in a number of papers; to be specific we shall concentrate here
on the description by Maiani {\it et al.}~\cite{Maiani:2015vwa}, in which the two $P_c$ states
(also denoted by the symbols  $\mathbb {P}^+(3/2^-)$ and $\mathbb {P}^+(5/2^+)$ ) have the valence
structure diquark-diquark-antiquark, as shown schematically in Fig.~\ref{ali:fig8} below. The 
assumed assignments are~\cite{Maiani:2015vwa}: 
\begin{align}
P_c(4380)^+= \mathbb {P}^+(3/2^-) &= \left\{\bar c\, [cq]_{s=1} [q^\prime q^{\prime \prime}]_{s=1}, L=0\right\},\nonumber 
\\
P_c(4450)^+= \mathbb {P}^+(5/2^+)&=\left\{\bar c\, [cq]_{s=1} [q^\prime q^{\prime \prime}]_{s=0}, L=1\right\}.\nonumber 
\end{align}
\noindent The observed mass difference $P_c(4450)^+ -P_c(4380)^+ \simeq 70$ MeV is accounted for as follows:
The level spacing for $\Delta L=1$ is set using the light baryons $\Lambda(1405)- \Lambda(1116) \sim 290$~MeV.
 The light-light diquark $[q^\prime q^{\prime \prime}]$ spin-dependent mass difference ($\Delta S=1$) is 
determined from the diquark-quark interpretation of the charm baryons 
$[qq^{\prime}]_{s=1} - [qq^{\prime}]_{s=0} =\Sigma_c(2455) - \Lambda_c(2286) \simeq 170$~MeV. 
Thus, the orbital mass gap $\mathbb {P}^+(3/2^-) - \mathbb {P}^+(5/2^+) $ is thereby reduced to
$120$~MeV, in approximate agreement with the data.

\vspace*{-5mm}
\begin{figure}
\centerline{\includegraphics[width=9.2cm]{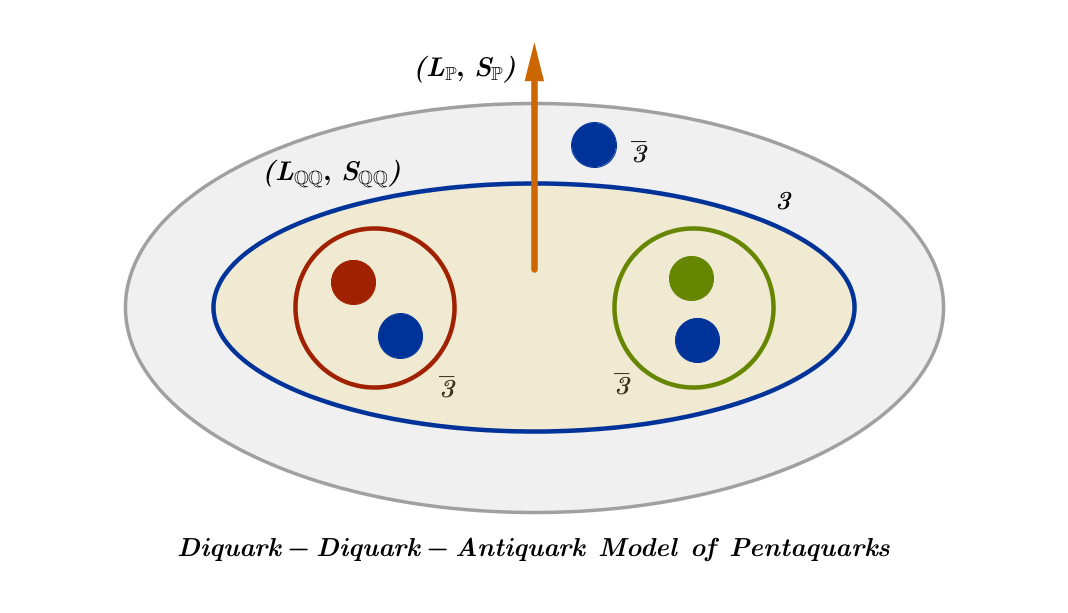}}
\vspace*{-4mm}
\caption{ }
\label{ali:fig8}
\end{figure}
\vspace*{-2mm}
Two possible mechanisms of pentaquark production in $\Lambda_b^0 \to K^- J/\psi p$
have been  proposed~\cite{Maiani:2015vwa}.
 In the first, the $b$-quark spin is shared between the $K^-$,
 the $\bar{c}$ and the $[cu]$~components, and the $[ud]$~diquark in the final state
retains its spin, i.e. it has  spin-0, (Fig.~\ref{ali:fig9}~A below). This is the decay mechanism compatible with
heavy-quark-spin conservation, which implies that the spin of the light diquark in $\Lambda_b^0$ decay is also
conserved. In the second, the $[ud]$~diquark is formed from the original $d$~quark, and the $u$~quark
from the vacuum $u\bar{u}$. In this case, angular momentum is shared among all components, and the diquark 
$[ud]$~may have both spins, $s=0, 1$ (Fig.~\ref{ali:fig9}~B below). 
Which of the two diagrams dominate is a dynamical question;  entries in the PDG on the decays
 of $\Lambda_b$~hint that the mechanism in Fig.~B is dynamically suppressed, as also anticipated by the
heavy-quark-spin conservation.

\begin{figure}
\centerline{\includegraphics[width=12cm]{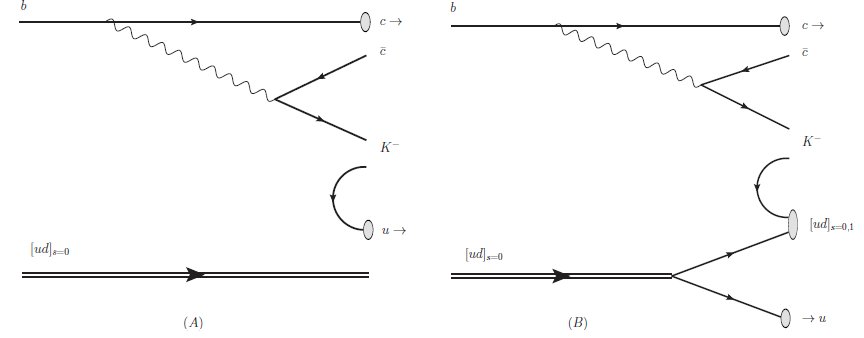}}
\caption{Two mechanisms for the decays $\Lambda_b^0 \to J/\psi  K^- p$ in the pentaquark picture~\cite{Maiani:2015vwa}.}  
\label{ali:fig9}
\end{figure}

\subsection{$SU(3)_F$ structure of pentaquarks}

Concentrating on the quark flavor of the pentaquarks $\mathbb {P}_c^+=\bar{c}cuud$, they are of two different types:
\begin{align}
\mathbb {P}_u &= \epsilon^{\alpha \beta \gamma}\bar c_\alpha\, [cu]_{\beta, s=0,1}\, [u d]_{\gamma,s=0,1},\nonumber 
\\
\mathbb {P}_d &= \epsilon^{\alpha \beta \gamma}\bar c_\alpha\, [cd]_{\beta, s=0,1}\, [u u]_{\gamma,s=1}\nonumber, 
\end{align}
the difference being that the $\mathbb {P}_d$ involves  a $[uu]$ diquark, and the Pauli exclusion principle
implies that this diquark has to be in an $SU(3)_F$-symmetric representation. 
This leads to two distinct $SU(3)_F$~series of pentaquarks
\begin{align}
\mathbb {P}_A &= \epsilon^{\alpha \beta \gamma}\left\{\bar c_{\alpha}\, [cq]_{\beta, s=0, 1}\,
 [q^\prime q^{\prime \prime}]_{\gamma, s=0}, L\right\}\nonumber 
= \mathbf { 3 \otimes \bar{3}= 1 \oplus 8},\nonumber\\
\mathbb {P}_S &= \epsilon^{\alpha \beta \gamma}\left\{\bar c_{\alpha}\, [cq]_{\beta, s=0, 1}\,
 [q^\prime q^{\prime \prime}]_{\gamma, s=1}, L\right\}\nonumber 
= \mathbf { 3 \otimes 6= 8 \oplus 10}.\nonumber
\end{align}
 For $S$~waves, the first and the second series have the angular momenta 
\begin{align}
\mathbb {P}_A(L=0) &:~~~J=1/2(2),~3/2(1),\nonumber\\
\mathbb {P}_S(L=0) &:~~~J=1/2(3),~3/2(3),~5/2(1),\nonumber
\end{align}
where the multiplicities are given in parentheses. One assigns $\mathbb {P}(3/2^-)$ to 
the $\mathbb {P}_A$ and $\mathbb{P}(5/2^+)$ to the $\mathbb {P}_S$~series of pentaquarks~\cite{Maiani:2015vwa}.

The $SU(3)_F$~based~analysis of the decays $\Lambda_b \to \mathbb {P}^+ K^- \to (J/\psi\, p) K^-$ goes as follows.
With respect to $SU(3)_F$, $\Lambda_b(bud)~\sim~\bar{3}$ and it is an isosinglet $I=0$. 
Thus, the weak non-leptonic Hamiltonian for $b\to c\bar{c}s$ decays is:
\begin{align}
H_{\rm eff}^{(3)} (\Delta I=0, \Delta S=-1).\nonumber 
\end{align}
The explicit form of the weak Hamiltonian is given by
\begin{align}
H_{\rm eff}^{(3)} = \frac{G_F}{\sqrt{2}} [ V_{cb} V_{cs}^* (c_1 O_1 + c_2 O_2)],\nonumber
\end{align}
where $G_F$ is the Fermi coupling constant, $V_{cs}$ is the CKM matrix element, $c_1$ and $c_2$ are the Wilson coefficients of the operators $O_1$
and $O_2$, respectively, with the operators defined as ($i,j$ are color indices)
\begin{align}
O_1= (\bar{s}_i c_j)_{V-A}(\bar{c}_i b_j)_{V-A},~~O_2= (\bar{s} c)_{V-A} (\bar{c} b)_{V-A}, \nonumber
\end{align}
and the penguin amplitudes are ignored.
With $M$ a nonet of $SU(3)$ light mesons $(\pi, K, \eta, \eta^\prime)$, the weak transitions  
$ \langle \mathbb {P}, M | H_{\rm W} |\Lambda_b\rangle$ requires $\mathbb{ P}+ M$  to be
in $8 \oplus 1$ representation.
Recalling the $SU(3)$ group multiplication rule
\begin{align}
8 \otimes 8 &= 1 \oplus 8 \oplus 8 \oplus 10 \oplus \bar{10} \oplus 27,\nonumber\\
8 \otimes 10 &= 8 \oplus 10 \oplus 27 \oplus 35,   \nonumber
\end{align}
the decay $ \langle \mathbb { P}, M| H_{\rm W} |\Lambda_b\rangle$ can be realized
 with $\mathbb {P}$ in
either an octet ${\tt 8}$ or a decuplet ${\tt 10} $.
The discovery channel $\Lambda_b \to \mathbb {P}^+ K^- \to J/\psi p K^-$ corresponds 
to $\mathbb {P}$ in an  octet ${\tt 8}$.
\subsection{Weak decays with $ \mathbb {P}$ in Decuplet representation}
 Decays involving the decuplet ${\tt 10} $ pentaquarks may also occur, if the 
light diquark pair having spin-0 $[ud]_{s=0}$ in $\Lambda_b$ gets broken to produce a spin-1~light
 diquark $[ud]_{s=1}$. In this case, one would also observe the decays of $\Lambda_b$, such as
\begin{align}
\Lambda_b \to \pi \mathbb {P}_{10}^{(S=-1)} & \to \pi(J/\psi \Sigma(1385)),\nonumber\\
\Lambda_b \to K^+ \mathbb {P}_{10}^{(S=-2)} & \to K^+ (J/\psi \Xi^-(1530)). \nonumber
\end{align}
These decays are, however, disfavored by the heavy-quark-spin-conservation selection rules.
The extent to which this rule is compatible with the existing data on $B$-meson and $\Lambda_b$ decays can
be seen in the PDG entries. Whether the decays of the pentaquarks are also subject to the same selection rules
is yet to be checked, but on symmetry grounds, we do expect it to hold. Hence, the observation (or not) of these
decays will be quite instructive.

 Apart from $\Lambda_b(bud)$}, several other $b$-baryons,
such as $\Xi_b^0(usb)$, $\Xi_b^-(dsb)$ and $\Omega_b^-(ssb)$ undergo weak decays. These $b$-baryons are characterized
by the spin of the light diquark, as shown below, making their isospin ($I$) and
strangeness ($S$) quantum numbers explicit as well as their light diquark $j^{\rm P}$ quantum numbers. 
\begin{figure}
\centerline{\includegraphics[width=0.8\textwidth]{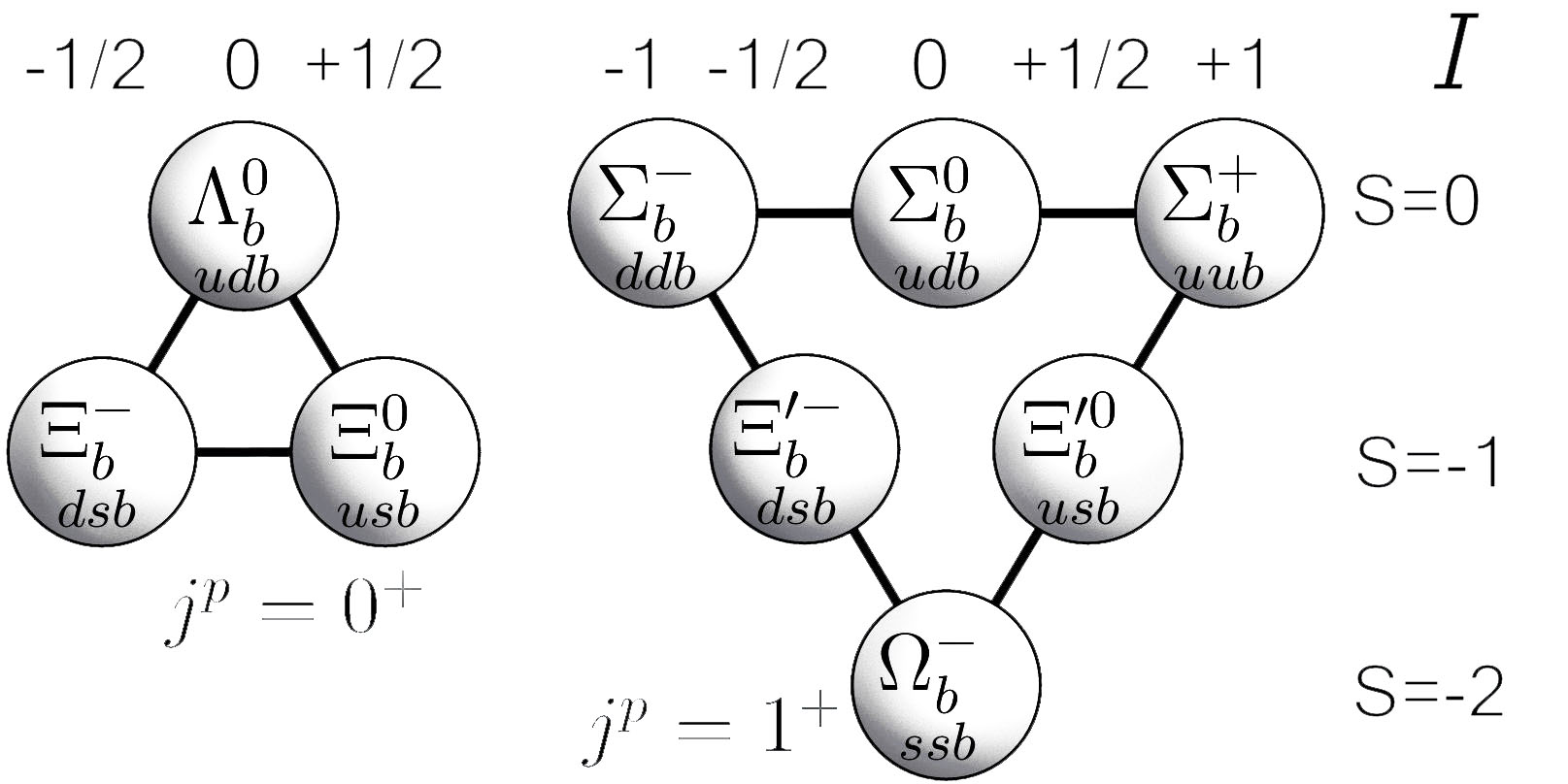}}
\caption{$b$-baryons with the light diquark  spins $j^p=0^+$ (left) and $j^p=1^+$ (right).}
\label{ali:fig10}
\end{figure}
The $c$-baryons are likewise characterized similarly.

\noindent Examples of bottom-strange b-baryon in various charge combinations,
 respecting $\Delta I=0,\, \Delta S=-1$ are:
 \vspace*{-2mm}
\begin{align}
\Xi_b^0(5794) & \to K (J/\psi \Sigma(1385)),\nonumber
\end{align}
which corresponds to the formation of the pentaquarks with the spin configuration
$ \mathbb {P}_{10} (\bar{c}\, [cq]_{s=0,1}\, [q^\prime s]_{s=0,1})$ with
 $(q,q^\prime=u,d)$.
%

\noindent  Above considerations have been extended involving the entire $SU(3)_F$ multiplets
entering the generic decay amplitude
$ \langle  {\cal P}  {\cal M}|H_{\rm eff} | {\cal B}\rangle,$
 where ${\cal B}$ is the 
$SU(3)_F$ antitriplet $b$-baryon, shown in the left frame of Fig.~\ref{ali:fig10},
${\cal M}$ is the $3\times 3$ pseudoscalar meson matrix\\
\begin{center}
$\qquad \qquad \mathcal{M}_{i}^{j}=\left( 
\begin{array}{ccc}
\frac{\pi ^{0}}{\sqrt{2}}+\frac{\eta _{8}}{\sqrt{6}} & \pi ^{+} & K^{+} \\ 
\pi ^{-} & -\frac{\pi ^{0}}{\sqrt{2}}+\frac{\eta _{8}}{\sqrt{6}} & K^{0} \\ 
K^{-} & \bar{K}^{0} & -\frac{2\eta _{8}}{\sqrt{6}}%
\end{array}%
\right)$,
\end{center}
and ${\cal P}$ is a pentaquark state belonging to an octet with definite
$J^P$, denoted as a $3\times 3$ matrix $J^P$, ${\cal P}^i_j(J^{\rm P})$,\\
\begin{center}
$\qquad \mathcal{P}_{i}^{j}\left( J^{P}\right) =\left( 
\begin{array}{ccc}
\frac{P_{\Sigma ^{0}}}{\sqrt{2}}+\frac{P_{\Lambda }}{\sqrt{6}} & P_{\Sigma
^{+}} & P_{p} \\ 
P_{\Sigma ^{-}} & -\frac{P_{\Sigma ^{0}}}{\sqrt{2}}+\frac{P_{\Lambda }}{%
\sqrt{6}} & P_{n} \\ 
P_{\Xi ^{-}} & P_{\Xi ^{0}} & -\frac{P_{\Lambda }}{\sqrt{6}}%
\end{array}%
\right) ,$\\
\end{center}
 or a decuplet ${\cal P}_{ijk}$ (symmetric in the indices), with
${\cal P}_{111}= \Delta^{++}_{10},...,{\cal P}_{333}=\Omega_{10}^-$.
(see Guan-Nan Li {\it et al.}~\cite{Li:2015gta} for a detailed list of the component fields
and $SU(3)_F$-based relations among decay widths).
The two observed pentaquarks are denoted as $P_p(3/2^-)$ and $P_p(5/2^+)$.
Estimates of the $SU(3)$ amplitudes require a dynamical model, which will be
lot more complex to develop than the factorization-based models for the two-body $B$-meson
decays, but, as argued in the literature, SU(3) symmetry can be used to relate
different decay modes.

Examples of the weak decays in which the initial $b$-baryon has a spin-1 light diquark, i.e. $j^{\rm P}=1^+$,
which is retained in the transition, are provided by the $\Omega_b$ decays.
The $s\bar{s}$ pair in $\Omega_b$ is in the symmetric ${\tt 6}$ representation
of $SU(3)_F$ with spin 1 and is expected to produce decuplet pentaquarks in association
with a $\phi$ or a Kaon~\cite{Maiani:2015vwa} 
\begin{align}
\Omega_b(6049) & \to \phi (J/\psi\, \Omega^-(1672)), K (J/\psi\, \Xi(1387)).\nonumber
\end{align}
These correspond, respectively, to the formation of the following pentaquarks ($q=u,\, d $)
\begin{align}
&\mathbb {P}^-_{10} (\bar{c}\, [cs]_{s=0,1}\, [ss]_{s=1}),
\mathbb {P}_{10} (\bar{c}\, [cq]_{s=0,1}\, [ss]_{s=1}).\nonumber
\end{align}
These transitions are expected on firmer theoretical footings, as the initial $[ss]$ diquark
in $\Omega_b$~is left unbroken. Again, lot  more transitions can be found relaxing this condition, which
would involve a $j^{\rm P}=1^+ \to 0^+$ light diquark, but they are anticipated to be suppressed. 

In summary, with the discoveries of the $X,Y,Z$ and $P_c$ states a new era of hadron spectroscopy 
has dawned. In addition to the well-known  $q\bar{q}$ mesons and $qqq$ baryons,
we have convincing evidence that the hadronic world is multi-layered, in the form of tetraquark mesons,
pentaquark baryons, and likely also the hexaquarks (or $H$ dibaryons)~\cite{Maiani:2015iaa}.
 However, the underlying dynamics is far from being understood.
 It has taken almost fifty years since the advent of QCD to develop quantitative 
understanding of the conventional hadronic physics.  A very long road lies
ahead of us before we can realistically expect to achieve a comparable understanding of multiquark hadrons.
Existence proof of tetra- and pentaquarks on the lattice would be a breakthrough.
In the meanwhile, phenomenological models built within constrained theoretical frameworks are 
unavoidable. They and experiments will guide us how to navigate through this uncharted territory.
The case of diquark models in this context was reviewed here. More data and  poweful theoretical 
techniques are needed to make further progress on this front. 

\section{Acknowledgements}
I would like to thank the organizers of the 14th. regional meeting on mathematical physics in Islamabad,
in particular, M. Jamil Aslam and Khalid Saifullah, for inviting me
and for their warm hospitality. Generous travel support offered by Dr. Shaukat Hameed Khan and  COMSTECH, Islamabad, is
also gratefully acknowledged.

\end{document}